# Phase front retrieval and correction of Bessel beams

B. MIAO[*], L. FEDER, J.E. SHROCK, AND H.M. MILCHBERG[*]

*Institute for Research in Electronics and Applied Physics, University of Maryland, College Park, MD 20742*
*[*]bmiao@umd.edu, [*]milch@umd.edu*

**Abstract:** Bessel beams generated with non-ideal axicons are affected by aberrations. We introduce a method to retrieve the complex amplitude of a Bessel beam from intensity measurements alone, and then use this information to correct the wavefront and intensity profile using a deformable mirror.

## 1. Introduction

A Bessel beam is a propagation-invariant solution of the Helmholtz equation where the transverse electric field distribution follows a $q^{th}$ order Bessel function, $E(r) = J_q(k_\perp r)$. Here, $k_\perp = |\mathbf{k}_\perp|$ is the perpendicular wave number, the total wavenumber is $\mathbf{k} = \mathbf{k}_\perp + \mathbf{k}_\parallel$, and $\mathbf{k}_\parallel = k\hat{\mathbf{z}}$ is the beam propagation wavenumber along $z$. Such an infinitely wide beam is "diffraction-free" [1] in the same trivial sense as a plane wave, but any practical realization of such a beam –also typically called a Bessel beam– has a finite aperture and thus is not diffraction-free. The axial extent of near-invariance along $z$, however, can be quite long by design, making these beams of high interest for applications which include optical trapping [2], laser machining [3,4], and optical coherence tomography [5]. Our particular interest has been the application of high intensity Bessel beams to generation of plasma waveguides for advanced laser-driven accelerators [6–11].

A Bessel beam can be viewed as resulting from the self-interference of a conical wave. It can be generated by imposing a conical phase shift on a Laguerre-Gauss beam using an axicon [12], a diffractive optical element [11,13], or a spatial light modulator (SLM) [14]. As the spatial frequency spectrum of a Bessel beam is an annulus in $\mathbf{k}_\perp$ space, it can also be generated with an annular aperture placed on the back focal plane of a converging lens [1].

For most applications, it is desired that generated Bessel beams have some minimum level of fidelity to the Bessel functional form, with a clear maximum on axis (for $J_0$) or off axis (for $J_{q>0}$), bounded by regions of near-zero field. This is a requirement for the generation of plasma waveguides, where the maximum intensity of Bessel beam must exceed the ionization threshold of the working gas [6,15]. In practice, Bessel beam fidelity is degraded by aberrations added by the phase-imposing element (axicons, diffractive optical elements, SLMs etc.) and in the incident beam, and so a straightforward, robust, and high-resolution method is needed for Bessel beam phase front correction. Prior work on Bessel beam astigmatism [16–19], due to either wavefront aberration or oblique illumination, has focused on the forward propagation problem rather than the phase retrieval problem.

While optical beam wavefronts can be directly characterized by interferometry or with Shack-Hartmann wavefront sensors [20,21], the drawbacks for a particular application may be setup complexity or intrinsic resolution limitations such as the number and size of micro-lenses in Shack-Hartmann sensors. In an indirect approach using a SLM, a beam's aberrated intensity profile is iteratively converged, via successive SLM-imposed phase corrections, to a reference profile such as a Gaussian. The corrected aberrations are the phase corrections accumulated over the optimization, which is achieved by the Gerchberg-Saxton algorithm [22], by complex modulation between orthogonal modes [23], by Stokes polarimetry [24], or by a least squares fitting algorithm using Zernike polynomials [25]. Indirect aberration correction of Bessel beams with a SLM has been reported in [26], using a beam propagation code.

For Bessel beam applications requiring high laser pulse intensities and large beam diameters, such as the generation of plasma waveguides [6–11], available SLMs are too small and have a relatively low optical damage threshold; reflective axicons [7,8] or thin diffractive optical elements [11] are the only choices for Bessel beam generation. For such applications, a method for Bessel beam wavefront correction based on intensity measurement alone is desirable. We present such a method using a deformable mirror (DM), and verify it both numerically and experimentally.

## 2. Bessel beam with aberrations

If $\tilde{E}(\rho, \theta, z=0) = \tilde{E}(\rho, \theta)$ is the electric field incident on a shallow axicon, forming a conical wavefront with base angle $\alpha \ll 1$, the diffraction integral, in the paraxial approximation, gives the complex field at $(r, \phi, z)$ as

$$U(r,\phi,z) = \frac{1}{i\lambda z}\exp\left(ikz + \frac{ikr^2}{2z}\right) \iint d\rho d\theta \rho \, \tilde{E}(\rho,\theta) \exp(ik\left(\frac{\rho^2}{2z} - \rho\tan\alpha - \frac{\rho r}{z}\cos(\phi-\theta)\right)). \quad (1)$$

Assuming $\tilde{E}(\rho, \theta)$ is a slowly varying function of $\rho$ and using the identity $e^{ix\cos\theta} = \sum_{n=-\infty}^{\infty} i^n J_n(x) e^{in\theta}$, we get

$$U(r,\phi,z) = \frac{1}{i\lambda z}\exp\left(ikz + \frac{ikr^2}{2z}\right)\sum_n (-i)^n e^{-in\phi} \int d\rho\rho \, \exp(ik\left(\frac{\rho^2}{2z} - \rho\tan\alpha\right))J_n(\frac{k\rho r}{z}) \int d\theta \, \tilde{E}(\rho,\theta) e^{in\theta}. \quad (2)$$

Equation (2) is further simplified using the method of stationary phase to give

$$U(r,\phi,z) = A(z)\sum_n (-i)^n c_n J_n(k_\perp r) e^{-in\phi}, \quad (3)$$

where $c_n = \int_0^{2\pi} \tilde{E}(\rho_z, \theta) e^{in\theta} d\theta$, $\rho_z = z\tan\alpha$, $k_\perp = k\tan\alpha$ and $A(z) = \sqrt{z/\lambda} \tan\alpha \exp(ik(z(1-\tan^2\alpha/2) + r^2/2z) - i\pi/4)$. Therefore, $\tilde{E}(\rho_z, \theta)$ maps to the field focal profile at $z$.

For generation of Bessel beams of order $q$ $(0, 1, 2, ...)$, we put $\tilde{E}(\rho_z, \theta) \to \tilde{E}(\rho_z, \theta)e^{iq\theta}$, where one can consider the incident field $\tilde{E}(\rho_z, \theta)$ as having passed through a spiral phase plate of order $q$ [7], so that $c_n = c_n^{(q)} = \int_0^{2\pi} \tilde{E}(\rho_z, \theta) e^{i(n+q)\theta} d\theta$. Therefore, Bessel beams can be decomposed into a series of orthogonal functions $J_n(k_\perp r) e^{-in\phi}$, with aberrations in the focal profile at $z$ determined by the phase distortion of the input beam in the annulus $(\rho_z, \theta)$. The phase retrieval problem reduces to a complex optimization problem, i.e. finding a set of coefficients

$$\{c_n^{(q)}\} = \arg\min \eta(\{c_n^{(q)}\}), \quad (4)$$

where $\eta(\{c_n^{(q)}\}) = \int (|\sum_n (-i)^n c_n^{(q)} J_n(k_\perp r) e^{-in\phi}|^2 - I_{M,z}(r,\phi))^2/I_{M,z}^2 \, rdrd\phi$ is the cost function and $I_{M,z}(r,\phi)$ is measured intensity profile at $z$. The complex minimization is completed using the nonlinear conjugate gradient method with Tensorlab [27,28]. To find the corrected transverse wavenumber $k_\perp^{opt}$, we first scan $k_\perp$ to minimize the cost function $\eta$ for $n=0$ only. Then $\eta(\{c_n^{(q)}\})$ is minimized with $|n| \leq 30$ for the examples in this paper. Once $\{c_n^{(q)}\}$ is found, this determines $U(r,\phi,z)$ and $\tilde{E}(\rho_z, \theta) = \sum_n \frac{1}{2\pi}\int_0^{2\pi} c_n^{(q)} e^{-i(n+q)\theta} d\theta$. We note that the minimization in Eq. (4) always converges to either $\tilde{E}(\rho,\theta)e^{iq\theta}$ or $\tilde{E}^*(\rho, \theta+\pi)e^{-iq(\theta+\pi)}$, corresponding to an output field of $U$ or $U^*$ respectively. Which of these results is applicable is resolved by comparing the extracted intensity profile $|\tilde{E}(\rho_z,\theta)|^2$ to the measured beam intensity profile. If the extracted profile is rotated by $\pi$ compared with the measured profile, then it corresponds to the complex conjugate solution. As we desire our Bessel beam

correction procedure to provide mainly rotational symmetry, aberrations with $0^{th}$ or $1^{st}$ order azimuthal dependence in $\cos\theta$ (e.g. defocus, spherical aberration and coma) are not considered in the following experiments and simulations.

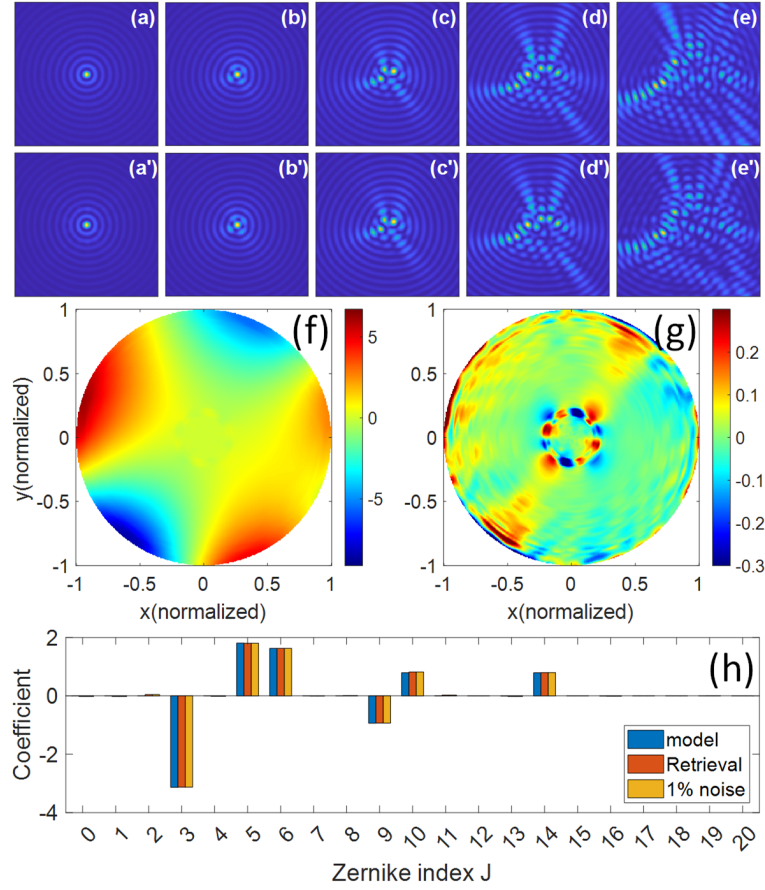

**Fig. 1.** Phase retrieval of synthesized Bessel beams with aberrations. (a)-(e) Synthesized $J_0$ at $z/\lambda = 2, 4, 6, 8$ and $10 \times 10^4$. (a')-(e') Corresponding retrieved $J_0$ with 1% noise added to (a)-(e). Each panel has width and height of 300 $\lambda$. (f) Retrieved phase map of the full $J_0$ input beam with 1% noise. The color bar shows phase in radians. (g) Difference between retrieved and input phase map (the retrieved and input maps are very close and look the same). (h) Zernike coefficients (up to J<21) of the input phase map, retrieved phase map, and retrieved phase map with 1% white noise added to input.

In Fig. 1, we demonstrate phase retrieval of a $J_0$ beam with aberrations up to the $4^{th}$ azimuthal order (from astigmatism to quadrafoil, or Zernike polynomials with ANSI index J<21[29]). The synthetic cases are calculated numerically from the non-paraxial diffraction integral

$$U(r,\phi,z) = \frac{1}{i\lambda z}\exp\left(ikz + \frac{ikr^2}{2z}\right) \iint d\rho d\theta \rho \, \tilde{E}(\rho,\theta) \exp(ikL - ik\rho \tan\alpha), \quad (5)$$

where $L = [z^2 + r^2 + \rho^2 - 2\rho r\cos(\theta - \phi)]^{1/2}$ and $\alpha = 0.05$ is the base angle of the conical wavefront (or the angle of the Bessel beam rays with respect to the beam axis). The grid size in the calculation is 20000$(r)\times$400$(\phi)$, with $\Delta r = 0.25\lambda, \Delta\theta = \pi/200$, where $\lambda$ is the wavelength. The input electric field is set as a flat-top, $\tilde{E}(\rho,\theta) = e^{i\psi(\rho,\theta)}$, with a wavefront phase aberration $\psi(\rho,\theta) = 5\,(\rho/\rho_{max})^2 \cos(2\theta + \pi/3) + 3\,(\rho/\rho_{max})^3 \sin(3\theta - \pi/6) + 2\,(\rho/\rho_{max})^4 \cos(4\theta - \pi/4)$ and a maximum beam radius $\rho_{max} = 5000\lambda$.

Figure 1 panels (a)-(e) show the intensity of the synthesized $J_0$ Bessel beam at axial positions $z/\lambda = 2, 4, 6, 8$ and $10 \times 10^4$, with random noise up to 1% of the peak intensity added in each frame. Figure 1 panels (a')-(e') show the corresponding retrieved $J_0$ Bessel beams using up to 30 modes ($n \leq 30$). The number of modes used in an intensity profile retrieval is increased until the cost function $\eta$ no longer decreases or it reaches a preset threshold (taken as 0.05 in these results).

The retrieved phase $\psi_{retr}(\rho, \theta)$ over the full input aperture is assembled from retrievals over annuli $0 \leq \rho_z \leq \rho_{max}$ and plotted in Fig. 1(f). The difference between the retrieved and input phases $\psi_{retr}(\rho, \theta) - \psi(\rho, \theta)$ is plotted in Figure 1(g), with the small error indicating excellent agreement. The larger error near the beam axis is possibly due to the inaccuracy of the paraxial approximation. For Fig. 1(h), we decompose the wavefront phase $\psi(\rho, \theta)$ into Zernike polynomials up to quadrafoil (ANSI index J< 21) and plot the coefficients for the cases of the input beam and retrieved beams with and without 1% noise. The retrieval is clearly robust against noise.

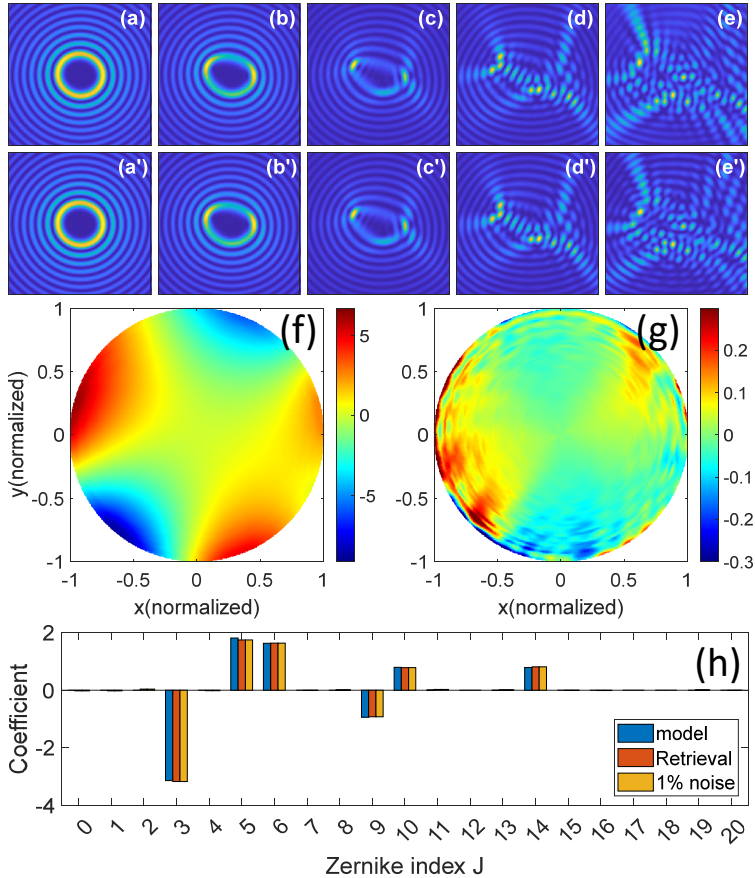

**Fig. 2** Phase retrieval of synthesized Bessel beams with aberrations. (a)-(e) Synthesized $J_8$ at $z/\lambda = 2, 4, 6, 8$ and $10 \times 10^4$. (a')-(e') Corresponding retrieved $J_8$ with 1% noise added to (a)-(e). Each panel has width and height of 300 $\lambda$. (f) Retrieved phase map of the full $J_8$ input beam with 1% noise. The color bar shows phase in radians. (g) Difference between retrieved and input phase map (the retrieved and input maps are very close and look the same). (h) Zernike coefficients (up to J<21) of the input phase map, retrieved phase map, and retrieved phase map with 1% white noise added to input.

Phase retrieval also works well for high order Bessel beams, as seen in Fig. 2. Here we impose the same aberrated phase $\psi(\rho, \theta)$ as earlier and take $\tilde{E}(\rho, \theta) = e^{i(\psi(\rho,\theta) + q\theta)}$, with $q = 8$. Figure 2(a)-(e) show the resulting synthesized high order $J_8$ Bessel beam for $z/\lambda =$

2, 4, 6, 8 and 10 × 10⁴, and panels (a′)-(e′) show the corresponding retrieved $J_8$ Bessel beam with up to 30 modes ($n \leq 30$). The retrieved wavefront $\psi_{retr}(\rho,\theta)$ is plotted in Fig. 2(f) and the error $\psi_{retr}(\rho,\theta) - \psi(\rho,\theta)$ is plotted in Fig. 2(g). Overall, the standard deviation of this error is less than 0.1 rad.

## 3. Experiment

We applied our method to characterize and correct Bessel beams with a reflective axicon. As shown in Fig. 3, a laser pulse (FWHM pulsewidth $\tau$=10 ns and diameter 4cm, $\lambda_0$=800nm) focused by a gold coated reflective axicon with $\alpha = 0.05$ base angle forms a $J_0$ Bessel beam. While the pulse is usually compressed to $\tau = 75$ fs for our plasma waveguide experiments, it is sufficient (and safer for the optics) to perform aberration corrections on the $\tau$=10 ns uncompressed pulse. The Bessel intensity profile $I_{M,z}(r,\phi)$ at $z$ is measured by the partial reflection from a plate beamsplitter through a 4mm diameter hole at the center of the reflective axicon. The beamsplitter is scanned along the entire focal line to extract the wavefront aberration $\psi_{retr}(\rho,\theta)$ over the full beam, as discussed above. Phase front corrections to the pre-compressor beam are provided by a deformable mirror (DM). Leakage of the beam through mirror M1 is sent to a shearing interferometry wavefront sensor (SID4, Phasics) to characterize the DM response. The modified Bessel focus is subsequently analyzed, and new correction terms added again to the DM, forming a closed loop and iteratively optimizing the Bessel focus. As the correction concerns only the rotational symmetry of the Bessel focus, as mentioned earlier, spherical and coma aberrations are not considered here.

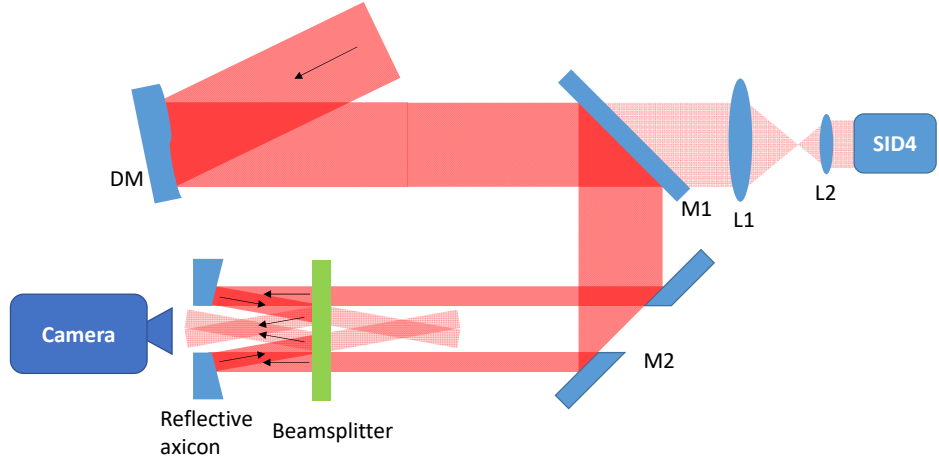

**Fig. 3.** Experimental setup for Bessel beam correction. DM: deformable mirror. SID4: shearing interferometry wavefront sensor. M1: Dielectric mirror. M2: Drilled dielectric mirror with 1-cm diameter hole. L1, L2: convex lenses.

Since the deformable mirror has a limited number of modes (eigenmodes of the mirror transfer matrix), we only corrected astigmatism and trefoil, which are due to axicon tilt and mechanical stress from the axicon mount. In Fig. 4, we show a sequence of measured ((a) to (e)) and retrieved ((a′) to (e′)) $J_0$ Bessel beam focal profiles evenly spaced along $z$ with no voltage on the deformable mirror. The cost function $\eta$ in these cases is no less than ~0.1, limited by single-shot image noise. The aberrated phase $\psi(\rho,\theta)$ over the full beam aperture is assembled from $\psi(\rho_z,\theta)$ measured at different $z$ positions and plotted in Fig. 4(f). The central part of the plot is missing due to mechanical constraints of the optical mounts and the central hole of the input beam. We decompose $\psi(\rho,\theta)$ into normalized Zernike polynomials in Fig. 4(d), with dominant terms J=3,5,6 and 9, which correspond to vertical/oblique astigmatism and vertical/oblique trefoil.

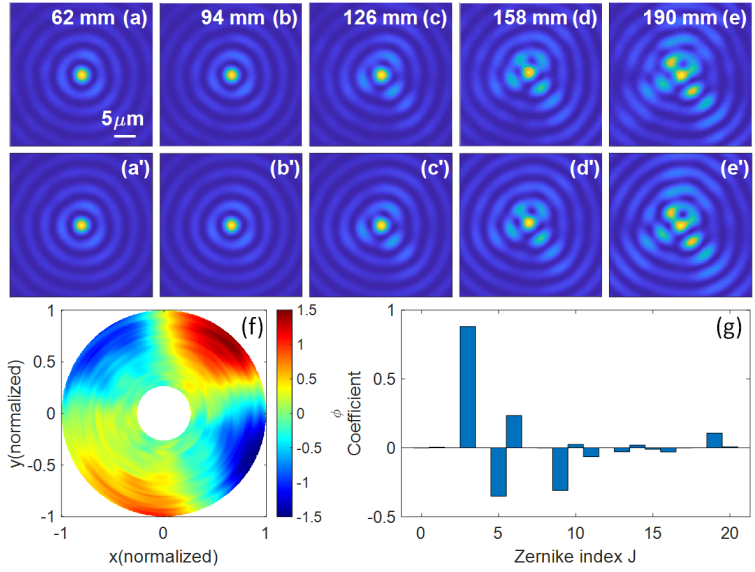

**Fig. 4.** Bessel beams with aberrations. (a)-(e) Measured focal intensity profiles at z= 62, 94, 126, 158 and 190 mm. (a′)-(e′) Retrieved focal intensity profiles at the same $z$ locations. (f) Reconstructed wavefront aberration $\psi$ over the full beam aperture (g) Decomposition of wavefront aberration into Zernike polynomials.

In practice, we found that correcting the Bessel beam profile only at the end of the axicon focus (see Fig. 4(e), $z = 190$ mm) resulted in a good correction for the whole focal line (See Fig. 5 and later discussion). This is because the surface aberration on the reflective axicons we measured was typically most significant near the outer edge of the axicon aperture, attributable to manufacturing error, or mechanical stress from mounting. Therefore, the wavefront aberration can be described by primary terms ($\rho^m \cos(m\theta)$ or $\rho^m \sin(m\theta)$, $m > 1$) and phase retrieval at the end of the focus is sufficient to measure the major contribution of the wavefront aberration. This can be seen in Fig. 4(g) from the dependence of the Zernike polynomials. For an arbitrary phase aberration, a measurement at a particular $z$ only gives the wavefront aberration on a particular annulus of the axicon. And because the aberration's radial distribution does not necessarily follow a single Zernike polynomial, correcting the aberration on a single annulus will only correct the focus at the corresponding $z$ location, not along the whole focal line. For correction over the full axicon aperture, phase retrieval needs to be carried out in a closed-loop along the full focal line and the correction involves higher order Zernike polynomials. Ultimately, the correction is limited by the number of actuators on the DM.

Having corrected the Bessel beam intensity profile at $z = 190$ mm, the profiles were recorded at other $z$ locations as shown in Fig. 5. Before correction, the Bessel beam intensity profiles from Fig. 4(a)-(e) (reproduced in Fig. 5(a)-(e)) show astigmatism and trefoil. After correction (panels 5(a′)-(e′)), the profiles along the whole focal line are improved, showing a single intense on-axis maximum for the beam. Panels 5(a″)-(e″) show the numerically retrieved, corrected intensity profiles. In comparison to Fig. 4(f) and (g), the wavefront aberration and its corresponding Zernike coefficients of Fig. 5(f) and (g) are much smaller. The Strehl ratio $S$ of the system [30] is expressed as the ratio of the central intensities of the aberrated and diffraction-limited point spread function, $S = |\langle e^{i\psi}\rangle|^2 \approx 1 - \sigma_\psi^2$. Here $\sigma_\psi^2 = \langle\psi^2\rangle - \langle\psi\rangle^2$ is the variance of the aberration and $\langle\psi\rangle = \int |\tilde{E}(\rho,\theta)|\psi(\rho,\theta)\rho d\rho d\theta / \int |\tilde{E}(\rho,\theta)|\rho d\rho d\theta$. Consider a special case of uniform illumination $\tilde{E}(\rho,\theta) = e^{i\psi(\rho,\theta)} = \exp(iB(\rho/\rho_{max})^m \cos m\theta)$, where $B$ is a constant. The Strehl ratio at a particular $z$ position is then $S = J_0^2(B(\rho_z/\rho_{max})^m)$, following Eq. (3). For the worst case, typically corresponding to the edge of the aperture, $S = J_0^2(B)$. The Strehl ratios of the Bessel beam at $z = 190$ mm before and after correction are $S = 0.43$ and $S = 0.98$, for a 2.3 × increase in focus intensity.

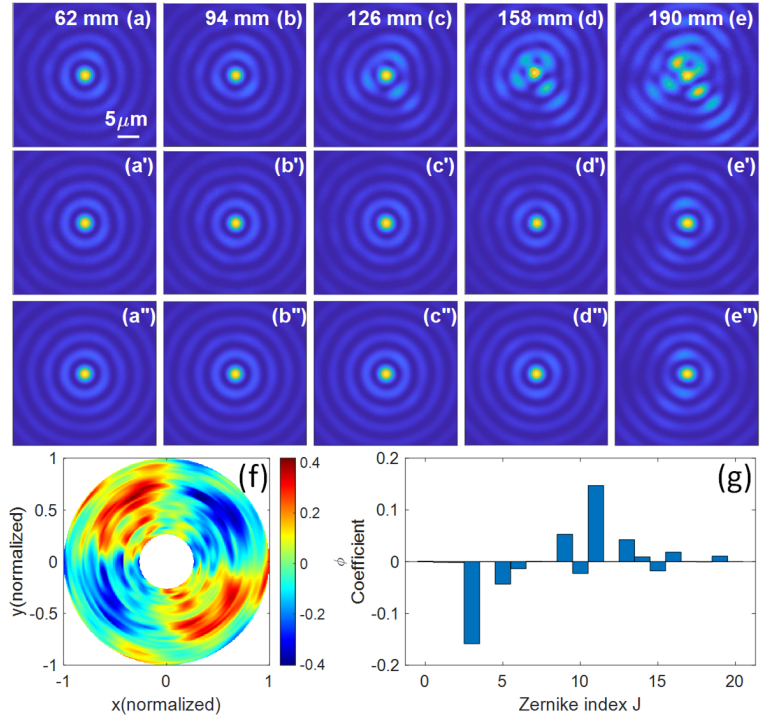

**Fig. 5.** Bessel focal profile optimized using a deformable mirror (DM). (a)-(e) Measured focal profiles at z=62,94,126,158 and 190 mm without correction by the DM. (a′)-(e′) Measured focal profiles at the same z locations with correction by the DM. (a″)-(e″) Numerically retrieved, corrected focal profiles at the same z locations. (f) Reconstructed wavefront aberration $\psi$ over the full beam aperture (g) Decomposition of wavefront aberration into Zernike polynomials.

In conclusion, we have presented a method for correcting aberrations in Bessel beam profiles using phase information extracted from intensity profile measurements alone. The method is validated both numerically and experimentally, demonstrating significant improvement of Bessel beam intensity profiles. The method is based on the extraction of complex aberration coefficients from nonlinear fitting of an analytic expression for an aberrated Bessel beam (Eq. (3)) to the measured profiles. We have found that for the specific axicons tested, and for our application of Bessel beams to plasma waveguide generation, only one intensity profile measurement was needed, at the $z$-location where the profile was most distorted, to collect the needed correction information.

**Funding.** US Department of Energy (DESC0015516) and the National Science Foundation (PHY1619582 and PHY2010511).

**Disclosures.** The authors declare no conflicts of interest.

**Data availability.** Data underlying the results presented in this paper are not publicly available at this time but may be obtained from the authors upon reasonable request.